\documentclass[12pt, leqno]{article}
\usepackage{amsmath,setspace,multirow,lineno}
\usepackage[font=small,labelfont=bf]{caption}
\usepackage[top=1.5in, bottom=1.5in, left=1.1in, right=1.1in]{geometry}

\usepackage[round]{natbib}
\usepackage{color,soul}

\DeclareCaptionStyle{italic}[justification=centering]{labelfont={bf},textfont={it},labelsep=colon}
\usepackage{graphicx,psfrag,epsf,textcomp,epstopdf, amsthm, paralist}
\usepackage{enumerate, dsfont, alltt, verbatim}
\usepackage{natbib}
\usepackage{url,xcolor}
\usepackage{booktabs, subfig, bm, paralist,mathpazo,tikz,todonotes,longtable,microtype, authblk}
\usepackage[linesnumbered,ruled,vlined]{algorithm2e}

\usepackage[pdftex,colorlinks=true]{hyperref}
\definecolor{darkblue}{rgb}{0,0,.6}
\hypersetup{citecolor=darkblue,linkcolor=darkblue,urlcolor=darkblue}
\definecolor{DarkRed}{rgb}{.7,0,.4}

\usepackage{comment,orcidlink}

\newcommand{\blind}{1}

\newcommand{\X}{\mathcal{X}}

\newcommand{\Rlogo}{\protect\includegraphics[height=1.8ex,keepaspectratio]{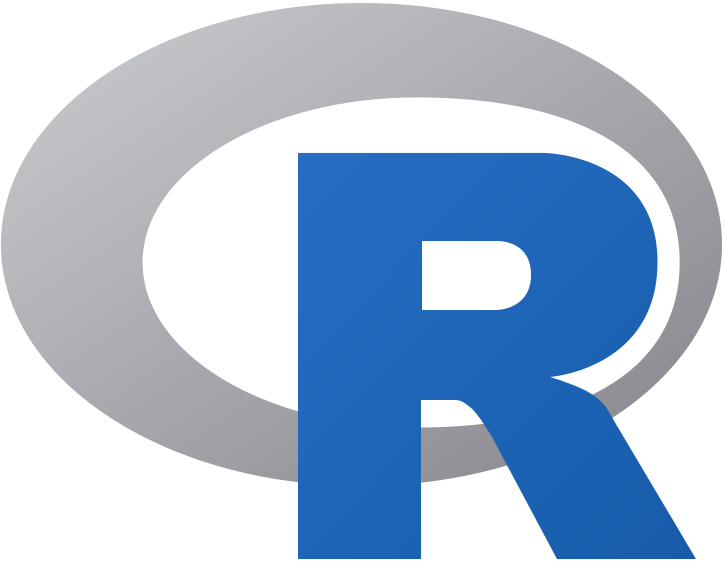}}

\graphicspath{{plots/}}

\newsavebox\CBox

 \newtheorem{@definition}{\sc Definition}[section]

  \renewcommand\X{\mathcal{X}}

\date{}

\begin{document}

\def\spacingset#1{\renewcommand{\baselinestretch}{#1}\small\normalsize} \spacingset{1}

\if1\blind
{
\title{\bf Functional Linear Cox Regression Model with Frailty}}
\author[1]{\normalsize Deniz Inan\orcidlink{0000-0002-0408-1309}}
\author[1]{\normalsize Ufuk Beyaztas \orcidlink{0000-0002-5208-4950}}
\author[2]{\normalsize Carmen D. Tekwe \orcidlink{0000-0002-1857-2416}}
\author[2]{\normalsize Xiwei Chen \orcidlink{0000-0003-1868-8387}}
\author[2]{\normalsize Roger S. Zoh\thanks{Corresponding address: School of Public Health-Bloomington 1025 E. Seventh Street, Suite 111
Bloomington, IN US.; Email: rszoh@iu.edu} \orcidlink{0000-0002-8066-1153}}

\affil[1]{\normalsize Department of Statistics, Marmara University, Turkey}
\affil[2]{\normalsize Department of Epidemiology and Biostatistics, Indiana University, School of Public Health, USA}
\maketitle
\fi

\begin{abstract}
In this paper, we introduce a functional linear Cox regression model with frailty to address unobserved heterogeneity in survival data involving functional covariates. Although conventional Cox models and their functional extensions are widely used, they are limited in handling frailty effects or random components for capturing individual differences not explained by observed covariates. Our proposed model integrates both scalar and functional covariates with a frailty term to account for unmeasured influences, thereby providing a robust framework for survival analysis in high-dimensional settings. We derive parameter estimates using functional principal component analysis and employ penalized partial likelihood to accommodate the frailty structure. A comprehensive simulation study demonstrates the model’s superior performance in terms of estimation accuracy and predictive capability over traditional functional linear Cox regression models and penalized functional regression models, especially under high-frailty conditions. In addition, we apply the proposed method to the analysis of a dataset from the National Health and Nutrition Examination Survey, revealing significant associations between physical activity patterns and mortality, even in frail subpopulations. Our results underscore the effectiveness of the proposed method in accurately capturing the complexity of survival data with both functional predictors and frailty, highlighting its potential applications in biomedical research where unobserved heterogeneity plays a critical role. Our method has been implemented in the \texttt{fcoxFr} \Rlogo \ package \footnote{The \Rlogo\ package \texttt{fcoxFr} is available at \url{https://github.com/denizlukusluinan/fcoxFr}}.
\end{abstract}

\noindent \textit{Keywords}: Cox regression, Functional principal component analysis, Frailty, Model selection. 

\newpage
\spacingset{1.65} 

\section{Introduction} \label{sec:1}

In survival analysis, the Cox proportional hazards model \citep{Cox1972} has been a cornerstone for analyzing time-to-event data due to its flexibility and semi-parametric nature. This model has seen extensive applications in various fields such as medicine, biology, and social sciences. However, traditional Cox models often fail to capture the inherent complexities in data where predictors involve curves or functions over a continuous domain \citep[see, e.g.,][]{Gellar2015}. To address this, the functional linear Cox regression model (FLCRM) has been developed, allowing the incorporation of functional predictors within this regression framework \citep[see, e.g.,][]{Clayton1978, Vaupel1979, McGilchrist1991, Terry2000, Duchateau2008}.

Let $T$ be the failure time, representing the time until the occurrence of an event of interest, such as death or relapse. Due to practical constraints, some subjects may not experience the event during the study period, resulting in censored observations. Let $C$ denote the censoring time, which is the time until the observation is censored (e.g., end of the study or loss to follow-up). The observed time $\widetilde{T}$ is then defined as $\widetilde{T} = \min(T, C)$, with an event indicator $\delta = I(T < C)$, where $I(\cdot)$ is the indicator function.

Consider a vector of scalar covariates $\bm{Z} \in \mathbb{R}^p$ consisting of $p$ individual components, specifically, $\bm{Z} = (Z_1, \ldots, Z_p)^\top$, and a functional predictor $\X(s)$ defined over a continuous domain $s \in \mathcal{I}$. The hazard function, denoted by $h(t \vert \bm{Z}, \X)$, for an individual with scalar covariates $\bm{Z}$ and a functional predictor $\X$ is expressed as the limit of the probability that the individual experiences an event within a short interval $[t, t + \bigtriangleup t)$ given survival up to time $t$, formulated as follows:
\begin{equation}\label{eq:hazard}
h(t \vert \bm{Z}, \X) = \lim_{\bigtriangleup t \rightarrow 0} \frac{P(t \leq T \leq t + \bigtriangleup t \vert T \geq t, \bm{Z}, \X)}{\bigtriangleup t}.
\end{equation}
Subsequently, the FLCRM that characterizes the hazard function is defined by:
\begin{equation}\label{eq:flcrm}
h(t \vert \bm{Z}, \X) = h_0(t) \exp \left(\bm{Z}^\top \bm{\gamma} + \int_{\mathcal{I}} \X(s) \beta(s) ds \right),
\end{equation}
where $h_0(t)$ is the baseline hazard function, representing the hazard for a subject with baseline covariates, $\bm{\gamma} = (\gamma_1, \ldots, \gamma_p)^\top$ is the vector of regression coefficients for the scalar covariates $\bm{Z}$, and $\beta(s)$ is the functional coefficient representing the effect of the functional predictor $\X(s)$ at location $s$. 

The functional parameter $\beta(s)$ is more complex to interpret compared to the non-functional parameter $\bm{\gamma}$. One way to understand it is that the term $\exp \left( \int_{\mathcal{I}} \beta(s) ds \right)$ represents the multiplicative increase in the hazard of death if the entire covariate function $\X(s)$ is increased by 1 unit, while keeping $\bm{Z}$ constant. In this scenario, $\beta(s)$ can be viewed as a functional log hazard ratio. More broadly, $\beta(s)$ acts as a weighting function for $\X(s)$, determining its overall impact on the hazard of mortality. The model assumes proportional hazards, meaning that this impact remains constant across the entire time domain $t$ of the hazard function \citep[see, e.g.,][for more discussion about the FLCRM]{Gellar2015}.

The primary objective of the FLCRM is to estimate the model parameters $[\bm{\gamma}, \beta(s)]$. Various methodologies have been introduced for this purpose. For instance, \cite{Gellar2015} and \cite{Qu2016} suggested maximizing penalized partial likelihood (PPL) functions to estimate the parameters in the model given by equation~\eqref{eq:flcrm}. Furthermore, \cite{Lee2015} proposed a Bayesian framework for parameter estimation within the same model. \cite{Kong2018} proposed a functional principal component (FPC) analysis-based approach to estimate the FLCRM. Furthermore, \cite{Yang2021} introduced a weighted version of the FLCRM and devised a three-stage estimation procedure to determine the model parameters. Recently, \cite{Crainiceanu2024} proposed a penalized functional regression (pfr) model to estimate the Model~\eqref{eq:flcrm} \citep[also see the \Rlogo \ package \texttt{refund}][for practical implication of the pfr method]{refund}.

All the aforementioned approaches were applied to complex empirical datasets, such as the Alzheimer’s Disease Neuroimaging Initiative (ADNI) and the National Health and Nutrition Examination Survey (NHANES) datasets, which present unique challenges for analyzing survival outcomes due to its high-dimensional and longitudinal nature. One of the primary obstacles is effectively integrating scalar covariates with functional predictors, such as Monitor-Independent Movement Summary (MIMS) and/or neuroimaging measurements over time, to predict survival times and disease progression. Traditional Cox regression models are inadequate for this task, as they do not account for the complexity and temporal dynamics inherent in functional predictors. The methodologies discussed successfully address these challenges by extending the Cox proportional hazards model to include functional predictors. By incorporating both scalar and functional predictors into the hazard function, the FLCRM provides a more accurate and nuanced understanding of the relationship between neuroimaging measurements and survival outcomes.

Despite its advances, current FLCRMs overlook a critical component often present in survival data: frailty. Frailty, an unobserved random effect, accounts for heterogeneity among subjects that is not captured by observed covariates. It introduces an additional layer of variability, allowing for more accurate modeling of survival times by accounting for individual susceptibility to the event of interest. Incorporating frailty into survival models is particularly important in biomedical research, where unmeasured genetic or environmental factors can significantly influence patient outcomes. Ignoring frailty can lead to biased parameter estimates and incorrect conclusions about covariate effects.

Frailty models offer a unique framework for addressing survival analysis problems, with two primary variations: unshared and shared frailty models. In the unshared model, each individual in the dataset is assumed to have their own unique random effect. In contrast, the shared frailty model is based on the assumption that individuals who share a common characteristic, such as siblings with the same mother or patients with a similar medical condition, also share a common random effect.

To date, the integration of frailty into FLCRMs has not been explored. Existing approaches to FLCRM focus primarily on the functional nature of the covariates, but assume a homogeneous population with respect to the risk of the event. This assumption is often unrealistic in practical applications. For example, in longitudinal studies of disease progression, patient responses to treatment and disease susceptibility can vary widely due to unmeasured factors, necessitating the inclusion of frailty for more accurate and robust modeling.

In this paper, we propose a novel extension to the FLCRM by incorporating a frailty term, thereby developing the functional linear Cox regression model with frailty (FLCRM-F). The inclusion of frailty in our model addresses the unobserved heterogeneity in survival data, providing a more comprehensive and realistic framework for analyzing functional covariates in the presence of unmeasured factors. Our approach leverages the strengths of both functional data analysis and frailty models, offering a significant improvement over existing methodologies.

We systematically develop the FLCRM-F, starting with the formulation of the model and estimation of parameters. We adopt an FPC analysis approach to handle the infinite-dimensional functional predictors by projecting them onto a finite-dimensional space. The frailty term is then incorporated into the Cox model, and we utilize appropriate estimation techniques to account for both the functional predictors and the frailty component. Our method is validated through extensive simulation studies and applied to real-world data from NHANES to demonstrate its practical utility and superior performance compared to existing FLCRMs.

In summary, the incorporation of frailty into FLCRMs represents a significant advancement in survival analysis with functional data. Our proposed FLCRM-F model not only fills a critical gap in the literature, but also offers a powerful tool for researchers dealing with complex survival data characterized by both functional predictors and unobserved heterogeneity.

The structure of this manuscript is outlined as follows: Section~\ref{sec:2} introduces the FLCRM-F model. In Section~\ref{sec:3}, we detail the estimation procedures for the proposed approach. Section~\ref{sec:4} presents a series of Monte Carlo experiments conducted to evaluate the method's performance in terms of estimation accuracy and predictive capability. Section~\ref{sec:5} discusses the application of our method to the NHANES dataset, with results highlighting its practical utility. Lastly, Section~\ref{sec:6} provides concluding remarks and explores potential directions to extend the methodology.

\section{Model and notations}\label{sec:2}

Let $\widetilde{\X} = \lbrace \widetilde{\X}(s) \rbrace_{s \in \mathcal{I}}$ represent a stochastic process satisfying $\int_{\mathcal{I}} \text{E}(\widetilde{\X}^2) < \infty$. The elements of this process, referred to as true trajectories, reside within the $\mathcal{L}^2$ Hilbert space, with $s$ confined to a bounded and closed interval $\mathcal{I} \subseteq \mathbb{R}$. Assuming without loss of generality, $\widetilde{\X}$ is a mean-zero stochastic process, i.e., $\text{E}[\widetilde{\X}(s)] = 0$. In practical applications, the functional curves or elements of $\widetilde{X}(s)$ are observed in a finite set of grid points $\lbrace s_j: j = 1, \ldots, J \rbrace \in \mathcal{I}$, where $J$ denotes the number of grid points. Consequently, the elements of $\widetilde{\X}(s)$ are measured with some error such that $\X(s_j) = \widetilde{\X}(s_j) + \varepsilon_j$, where $\varepsilon_j$ are independent and identically distributed random variables with a mean of zero and variance $\sigma^2$. 

Let $T$, $C$, and $\widetilde{T}$ represent the time to an event of interest, the censoring time, and the observed time, respectively. The observed time is defined as $\widetilde{T} = \min(T, C)$, with an event indicator $\delta = I(T < C)$. Furthermore, let $\bm{w}$ denote the vector of frailties or random effects representing unobserved heterogeneity among individuals, usually assumed to follow a distribution $\bm{w} \sim \Omega (\bm{w}; \bm{V}(\bm{\alpha}))$, with mean zero and covariance matrix $\bm{V}(\alpha)$, with $\bm{\alpha}$ denoting a vector of unknown parameters of the distribution $\Omega$. Consider a random sample $\lbrace \X_i(s), \bm{Z}_i, \widetilde{T}_i, \delta_i, \bm{w}_i: i = 1, \ldots, n \rbrace$ drawn from the joint distribution of $\lbrace \widetilde{\X}, \bm{\mathcal{Z}}, \widetilde{T}, \delta, \bm{w} \rbrace$, where $\bm{Z}_i = (Z_{1i}, \ldots, Z_{pi})^\top$ is a $p$-dimensional scalar covariate for the $i^\textsuperscript{th}$ individual. Then, the proposed FLCRM-F model, which characterizes the hazard function incorporating frailty, is defined as follows:
\begin{equation}\label{eq:frlmod}
h_i(t \vert \bm{Z}_i, \X_i, \bm{w}) = h_0(t) \exp \left(\bm{Z}_i^\top \bm{\gamma} + \int_{\mathcal{I}} \X_i(s) \beta(s) ds + \bm{U}_i^\top \bm{w} \right),
\end{equation}
where $h_0(t)$ is the baseline hazard function, $h_i(t \vert \bm{Z}_i, \X_i, \bm{w})$ is the hazard function incorporating frailty for subject $i$, $\bm{\gamma} = (\gamma_1, \ldots, \gamma_p )^\top$ is the vector of regression coefficients, $\beta(s)$ is the functional regression coefficient, and the frailty for subject $i$, $\bm{w}_i$, captures the effect of unobserved factors influencing the risk of the event for subject $i$. Here, $\bm{U}_i = (U_{1i}, \ldots, U_{di})^\top$ is a $d$-dimensional indicator vector that indicates the family of $i^\textsuperscript{th}$ individual. $\bm{U}_{ji} =1$ when $i^\textsuperscript{th}$ individual is a member of family $j$ and 0 otherwise. It is important to note that the matrix $\bm{U} = (\bm{U}_1, \ldots, \bm{U}_d)^\top$ becomes the identity matrix in the case of an unshared Cox regression model, where, $k = n$. Here, $d$, the number of random effects, is prespecified if the grouping structure is known or selected during estimation if the grouping structure is determined dynamically. In addition, $\bm{U}$ is full rank if the grouping structure is well-defined and all groups are represented without linear dependencies. Special cases like unshared frailty (identity matrix) guarantee full rank.

\section{Estimation}\label{sec:3}

This section outlines the estimation procedure developed in this study to determine the model parameters in~\eqref{eq:frlmod}, that is, $[\bm{\gamma}, \beta(s)]$. The infinite-dimensional nature of the regression coefficient function $\beta(s)$ poses a significant challenge for direct estimation. To address this, our procedure begins by projecting all functional components in~\eqref{eq:frlmod} onto a finite-dimensional space using FPC. This transformation converts the infinite-dimensional structure in~\eqref{eq:frlmod} into a finite-dimensional Cox regression model with frailty, utilizing FPC basis expansion coefficients.

Consider the covariance function $G(s, u) = \text{Cov}[\X(s) \X(u)]$. Using the Karhunen-Lo\'eve theorem, under appropriate regularity conditions, the covariance kernel $G(s, u)$ can be decomposed as:
\begin{equation*}
G(s, u) = \sum_{k=1}^{\infty} \lambda_k \phi_k(s) \phi_k(u), \quad \forall t,u \in \mathcal{I},  
\end{equation*}
where $\lbrace \phi_k(s): k = 1, 2, \ldots \rbrace$ are orthonormal eigenfunctions (FPCs) in the $\mathcal{L}^2$ sense corresponding to the non-negative eigenvalues $\lbrace \lambda_k: k = 1, 2, \ldots \rbrace$. It is assumed that the eigenvalues of $G(s, u)$ are distinct to ensure the uniqueness of the orthonormal bases of the eigenfunctions. Each realization of $\X_i(s)$ can then be expanded as:
\begin{equation*}
\X_i(s) = \sum_{k=1}^{\infty} \xi_{ik} \phi_k(s), \quad i = 1, \ldots, n,
\end{equation*}
where for a given $i$, the random variables $\xi_{ik} = \int_{\mathcal{I}} \X_i(s) \phi_k(s)$ represent the uncorrelated FPC scores with variance $\lambda_k$. In practice, most of the variability in functional variables can be captured by the first few eigenfunctions. Therefore, we project $\X_i(s)$ onto basis expansions with a truncation constant $K$. Consequently, each realization $\X_i(s)$ is approximated by:
\begin{equation*}
\X_i(s) \approx \sum_{k=1}^{K} \xi_{ik} \phi_k(s), \quad i = 1, \ldots, n,
\end{equation*}

Using the FPC decomposition, the integral part in the proposed FLCRM-F model in~\eqref{eq:frlmod} can be approximated as follows:
\begin{align}
\int_{\mathcal{I}} \X_i(s) \beta(s) ds &= \int_{\mathcal{I}} \sum_{k=1}^K \xi_{ik} \phi_k(s) \beta(s) ds, \nonumber \\
&= \sum_{k=1}^K \xi_{ik} \int_{\mathcal{I}} \phi_k(s) \beta(s) ds, \nonumber \\
&= \sum_{k=1}^K \xi_{ik} \beta_k, \label{eq:fpc}
\end{align}
where $\beta_k = \int_{\mathcal{I}} \beta(s) \phi_k(s) ds$ are the uncorrelated FPC scores of $\beta(s)$.

Substituting~\eqref{eq:fpc} into~\eqref{eq:frlmod}, we get
\begin{equation*}
h_i(t \vert \bm{Z}_i, \X_i, \bm{w}) = h_0(t) \exp \left(\bm{Z}_i^\top \bm{\gamma} + \sum_{k=1}^K \xi_{ik} \beta_k + \bm{U}_i^\top \bm{w}  \right).
\end{equation*}
Define $\bm{\theta} = (\bm{\gamma}, \bm{\beta})^\top$ with $\bm{\beta} = (\beta_1, \ldots, \beta_K)^\top$ and $\bm{D}_i = (\bm{Z}_i^\top, \bm{\Xi}_i)^\top$ with $\bm{\Xi}_i = (\xi_{i1}, \ldots, \xi_{iK})^\top$. Thus, the FLCRM-F model with the FPC approximation of the functional components can be written as:
\begin{equation}\label{eq:frlmod_d}
h_i(t \vert \bm{Z}_i, \X_i, \bm{w}) = h_0(t) \exp \left(\bm{D}^\top_i \bm{\theta} +  \bm{U}_i^\top \bm{w} \right).
\end{equation}
Using FPC decomposition, the estimation problem for the infinite-dimensional FLCRM-F model is reduced to estimating a finite-dimensional Cox regression model with frailty. In real-world scenarios, even though $\X_i(s)$ lies in an infinite-dimensional space, it is generally observed at specific discrete time points with some measurement error, as discussed in Section~\ref{sec:2}. To address this issue, functional forms of discretely observed data are typically approximated using basis expansion techniques such as B-splines or Fourier series. In this research, we employ the B-spline basis expansion method \citep{Eilers1996} due to its straightforward application and proven effectiveness.

Conditionally on $\bm{w}$, the likelihood for model~\eqref{eq:frlmod_d} is given as follows:
\begin{align*}
L(h_0(t), \bm{\theta}, \bm{\alpha}) &= \int \prod_{i=1}^{n} h_i(t \mid \bm{w})^{\delta_i} S_i(t \mid \bm{w}) \Omega(\bm{w}; V(\bm{\alpha})) \, d \bm{w} \\
&= \int \prod_{i=1}^{n} \left[ h_0(t) \exp(\bm{D}_i \bm{\theta} + \bm{U}_i^\top \bm{w}) \right]^{\delta_i}
\exp \left[ -H_0(t) \exp(\bm{D}_i \bm{\theta} + \bm{U}_i^\top w) \right]
\Omega(\bm{w}; V(\bm{\alpha})) \, d \bm{w},
\end{align*}
where $H_0(t)=\int_{0}^{t}h_0(u)du$ is the cumulative hazard function and $S_{i}(t \mid \bm{w}) = \exp(-\bm{w} H_{0}(t))$ ($ = \left[ -H_0(t) \exp(\bm{D}_i \bm{\theta} + \bm{U}_i^\top w) \right]$) conditional survival function for an individual. 

In this study, we assume that the frailty vector \( \bm{w} \) follows a multivariate normal distribution. However, the likelihood approximations derived in the subsequent sections can be readily extended to accommodate other frailty distributions \citep[see, e.g.,][]{ripatti2000estimation}. When the frailties are normally distributed, the integral becomes analytically intractable. To overcome this, the Laplace approximation method, as proposed by \cite{breslow1993approximate}, offers an effective way to approximate the marginal log-likelihood, as follows:
\begin{align}\label{eq:loglikelihood}
\ell((h_0(t), \bm{\theta}, \bm{\alpha}) &\approx - \frac{1}{2} \log \vert V(\bm{\alpha}) \vert \\\nonumber
&- \frac{1}{2} \log \vert \sum_{i=1}^{n} H_0(t) \exp(\bm{D}_i \bm{\theta} + \bm{U}_i^\top \bm{w} ) \bm{U}_i \bm{U}_i^\top - V(\bm{\alpha})^{-1} \vert \\ \nonumber
&+ \sum_{i=1}^{n} \delta_i \left[ \log (h_0(t)) + \bm{D}_i \bm{\theta} + \bm{U}_i^\top \bm{w} \right] \\ \nonumber
&- H_0(t) \exp(\bm{D}_i \bm{\theta} + \bm{U}_i^\top \bm{w}) - \frac{1}{2} \bm{w}^\top V(\bm{\alpha})^{-1} \bm{w}. \nonumber
\end{align}

Assuming $\bm{\alpha}$ is known and $\bm{w}$ is treated as a fixed effects parameter, the last two lines of~\eqref{eq:loglikelihood} constitute the full likelihood for a Cox model, with $\bm{w}$ representing an additional set of parameters. The expression $(\frac{1}{2} \bm{w}^\top V(\bm{\alpha})^{-1} \bm{w})$ acts as a penalty term, shrinking extreme values of $\bm{w}$. This likelihood can be maximized through the PPL approach as follows: 
\begin{align}
\sum_{i=1}^{n} \delta_i \left(( \bm{D}_i \bm{\theta} + \bm{U}_i^\top \bm{w} ) - \log \sum_{j \in R(t_i)} \exp \left( \bm{D}_j \bm{\theta} + \bm{U}_j^\top \bm{w}  \right) \right) 
- \frac{1}{2} \bm{w}^\top V(\bm{\alpha})^{-1} \bm{w}, \nonumber
\end{align}
where $R(t_i)$ represents the risk set at time $t_i$.

Given $\bm{\alpha}$, ignoring the first two lines of~\eqref{eq:loglikelihood} and using the PPL to estimate $(\hat{\bm{\theta}}(\bm{\alpha}), \hat{\bm{w}}(\bm{\alpha}))$, the estimating equations based on the first partial derivates of the PPL with respect to $\bm{\theta}$ and $\bm{w}$ are respectively:
\begin{align}\label{eq:estimatingteta}
\sum_{i=1}^{n} \delta_i \left[\bm{D}_i- \frac{\bm{D}_i \exp(\bm{D}_i \bm{\theta} + \bm{U}_i^\top \bm{w} )}{\sum_{j\in R(t_{i})}^{ } \exp(\bm{D}_j \bm{\theta} + \bm{U}_j^\top \bm{w} )}\right]=0,
\end{align}
and:
\begin{align}\label{eq:estimatingw}
\sum_{i=1}^{n} \delta_i \left[\bm{U}_i- \frac{\bm{U}_i \exp(\bm{D}_i \bm{\theta} + \bm{U}_i^\top \bm{w} )}{\sum_{j\in R(t_{i})}^{ } \exp(\bm{D}_j \bm{\theta} + \bm{U}_j^\top w )}\right]-V(\bm{\alpha})^{-1} \bm{w}=0
\end{align}

The parameters $\left( \hat{\theta}(\bm{\alpha}), \hat{\bm{w}}(\bm{\alpha}) \right)$ are determined by alternating between the solutions of ~\eqref{eq:estimatingteta} and ~\eqref{eq:estimatingw}. By plugging the maximized values $\left( \hat{\theta}(\bm{\alpha}), \hat{\bm{w}}(\bm{\alpha}) \right)$ obtained from the PPL into ~\eqref{eq:loglikelihood}, we derive an approximate profile likelihood function as a function of $\bm{\alpha}$ as follows:
\begin{align}\label{eq:profilelikelihood}
\ell(\hat{\theta}(\bm{\alpha}), \bm{\alpha}) &\approx - \frac{1}{2} \log \vert V(\bm{\alpha}) \vert\\\nonumber
& - \frac{1}{2} \log \vert \sum_{i=1}^{n} H_0(t) \exp(\bm{D}_i \hat{\theta}(\bm{\alpha}) + \bm{U}_i^\top \hat{\bm{w}} ) \bm{U}_i \bm{U}_i^\top - V(\bm{\alpha})^{-1} \vert - \frac{1}{2} \hat{\bm{w}}^\top V(\bm{\alpha})^{-1} \hat{\bm{w}}.
\end{align}

Denoting the $ \sum_{i=1}^{n} H_0(t) \exp(\bm{D}_i \hat{\theta}(\bm{\alpha}) + \bm{U}_i^\top \hat{\bm{w}} ) \bm{U}_i \bm{U}_i^\top - V(\bm{\alpha})^{-1}$ as $K(\hat{\bm{w}})$, ~\eqref{eq:profilelikelihood} can be written as follows:
\begin{align}\label{eq:reprofilelikelihood}
\ell(\hat{\theta}(\bm{\alpha}), \bm{\alpha}) &\approx - \frac{1}{2} \log \vert V(\bm{\alpha}) \vert\\\nonumber
& - \frac{1}{2} \log \vert \bm{K}(\hat{w})\vert - \frac{1}{2} \hat{\bm{w}}^\top V(\bm{\alpha})^{-1} \hat{\bm{w}}.
\end{align}
Setting the first derivative of ~\eqref{eq:reprofilelikelihood} with respect to $\bm{\alpha}$ to zero and subtituing $H_0(t)$ by its Nelson-Aalen \citep{aalen1978nonparametric} estimate, $\bm{\alpha}$ is estimated as follows:
\begin{align}\label{eq:estimatealpha}
\hat{\bm{\alpha}} = \frac{\hat{\bm{w}}^\top \hat{\bm{w}} + \text{tr}(K(\hat{\bm{w}})^{-1})}{k}.
\end{align}

The process of maximizing the approximate likelihood ~\eqref{eq:loglikelihood} is performed in two main steps. First, with an initial value of $\bm{\alpha}$, the proposed equations ~\eqref{eq:estimatingteta} and ~\eqref{eq:estimatingw} are solved using the Newton-Raphson algorithm. Once solutions for $\bm{\theta}$ and $\bm{w}$ are obtained, these values are fixed, and equation ~\eqref{eq:estimatealpha} is then used to update $\bm{\alpha}$. Subsequently, equations ~\eqref{eq:estimatingteta} and ~\eqref{eq:estimatingw} are recalculated, and the process is repeated iteratively until convergence is achieved \cite[please see,][for more information about the iterative algorithm]{ripatti2000estimation}. Upon convergence, the cumulative baseline hazard can be estimated using Breslow's estimator (\cite{breslow1974covariance}).

Let $\widehat{\bm{\theta}} = (\widehat{\bm{\gamma}}, \widehat{\bm{\beta}})^\top$, with $\bm{\beta} = (\widehat{\beta}_1, \ldots \widehat{\beta}_K)^\top$, obtained by PPL. Then, the estimate of the regression coefficient function $\beta(t)$ is obtained as follows:
\begin{equation*}
\widehat{\beta}(t) = \sum_{k=1}^K \widehat{\beta}_k \phi_k(t).
\end{equation*}

\section{Simulation study}\label{sec:4}

We conducted an extensive series of Monte Carlo experiments to assess the finite sample performance of the proposed method, FLCRM-F, and compare its finite sample performance with pfr and FLCRM under a modified version of the data generation process originally presented in \cite{Kong2018}. In the experiments, synthetic datasets are generated according to the model specified in~\eqref{eq:frlmod}, with $p = 6$ scalar covariates and a functional predictor. 

The functional predictor $\widetilde{\X}(s)$  is generated over the interval $[0,1]$ at 101 equidistant points using the following process:
\begin{equation*}
\widetilde{\X}_i(s) = u_{i1} + u_{i2} s + \sum_{j=1}^{10} [v_{ij1} \sin \{2 (2j - 1) \pi s\} + v_{ij2} \cos\{2 (2j -1) \pi s\}],
\end{equation*}
where $u_{i1}, u_{i2} \sim \mathcal{N}(0,1)$ and $v_{ij1}, v_{ij2} \sim \mathcal{N}(0, j^2)$. The scalar covariates $\bm{Z}_i = (Z_{i1}, \ldots, Z_{i6})^\top$ are sampled from a multivariate normal distribution with zero mean and covariance matrix $\bm{\Sigma} = (\rho^{\vert j-k \vert})_{1 \leq j, k \leq 6}$ with $\rho = 0.5$. Additionally, similarly to \cite{Kong2018}, we introduce correlation between the scalar covariates and the functional predictor by setting $\text{Cov}(z_{ik}, v_{i11}) = 0.1$ for $1 \leq k \leq 6$. To simulate noisy observations, we add a noise term $\epsilon_i(s)$ to the functional predictor, yielding $\Xi(s) = \widetilde{\X}_i(s) + \epsilon_i(s_m)$, where $\epsilon_i(s_m) \sim \mathcal{N}(0, 0.5)$ for $m = 1, \ldots, 101$. This setup allows for a realistic evaluation of the robustness and predictive capacity of the proposed method in the presence of measurement error. 

As discussed in Section~\ref{sec:3}, our proposed method estimates the model parameters assuming log-normal distributed frailty terms. Consequently, using frailty terms generated from a log-normal distribution would result in an ideal fit for our method. To ensure a fair comparison between our method, pfr, and the FLCRM, we generate the frailty term, $\bm{w}$, from a gamma distribution with shape and scale parameters $1/\varphi$ and $\varphi$, respectively, i.e., $\bm{w} \sim \text{Gamma}(1/\varphi, \varphi)$, where $\text{Var}(\bm{w}) = \varphi$. 

Scalar covariates are fixed at $\bm{\gamma} = [0.5, 0.5, 0.5, 0.5, 0.5, 0.5]^\top$, and the functional coefficient is defined as
\begin{align*}
\beta(s) &= 0.3 [\sin(\pi s) - \cos(\pi s) + \sin(3 \pi s/10) - \cos(3 \pi s) + \sin(5 \pi s)/9 - \cos(5 \pi s)/9 + \sin(7 \pi s)/16 \\
&- \cos(7 \pi s)/16 + \sin(9 \pi s)/25 + \cos(9 \pi s)/25 + (2 \pi)^{-1/2} \exp \{-2^{-1} (s-0.5)^2 \} ], \quad 0 \leq s \leq 1.
\end{align*}
The failure time $T_i$ for each subject is independently generated from an
exponential distribution with rate parameter $\exp \left(\bm{Z}_i^\top \bm{\gamma} + \int_0^1 \X_i(s) \beta(s) ds + \bm{U}_i^\top \bm{w} \right)$. The censoring times are independently drawn from an exponential distribution with the rate parameter $\tau$, i.e., $\mathcal{C}_i \sim \exp(\tau)$. Because we restrict our focus to unshared frailty case in our simulation studies, $\bm{U}$ is an identity matrix.

Throughout the simulation study, we consider four distinct sample sizes: $n = [100, 250, 500, 1000]$. To induce varying levels of censoring, four different rate parameters, $\tau = [0.01, 0.1, 0.2, 0.3]$, are used to generate censoring times, resulting in a range of censoring rates (approximately, the range is between 0 and 0.45 as presented in Table~\ref{tab:tab_1}). In addition, we examine the impact of frailty effects by considering four different values of the frailty variance, $\varphi = [0.01, 1, 1.5, 2]$, which controls the unobserved heterogeneity in the data. In scenarios where $\varphi = 0.01$,  the influence of unobserved frailty on the risk of the event for subject $i$ is minimal. Therefore, we expect that all the methods (i.e., pfr, FLCRM, and FLCRM-F) will demonstrate similar performance in terms of estimation accuracy. However, for larger values of $\varphi$, the frailty effect is more pronounced, and the FLCRM-F method is anticipated to outperform the pfr and standard FLCRM, given its ability to account for the increased heterogeneity.

For each dataset generated, we estimate the model parameters and evaluate the estimation performance using the following mean squared error (MSE) and integrated mean squared error (IMSE) metrics for the scalar coefficient vector $\bm{\gamma}$ and the functional coefficient $\beta(s)$, respectively:
\begin{equation*}
\text{MSE} = \Vert \widehat{\bm{\gamma}} - \bm{\gamma} \Vert^2, \qquad \text{IMSE} = \int_0^1 [\widehat{\beta}(s) - \beta(s)]^2 ds,
\end{equation*}
where $\widehat{\bm{\gamma}}$ and $\widehat{\beta}(s)$ are the estimates of $\bm{\gamma}$ and $\beta(s)$, respectively. Additionally, for each combination of $n$, $\varphi$, $\tau$, we compute the concordance index $\text{CI}_{in}$ on the training data to assess how well the models' predicted risk scores align with the observed event times. To further evaluate predictive capabilities of the models, an independent test sample of size $n$ is generated for each parameter configuration, and the out-of-sample concordance index $\text{CI}_{out}$ is computed on these test datasets. This allows for a thorough assessment of the models' discriminatory power. Furthermore, the observed mean censoring rate, denoted by $\psi$ is calculated for each parameter combination. To ensure robust evaluation, the entire simulation process is repeated 500 times, and the average performance metrics, including MSE, IMSE, $\text{CI}_{in}$, $\text{CI}_{out}$, and $\psi$,  are reported across all replications.

Our results are presented in Table~\ref{tab:tab_1}. The results consistently show that increasing the sample size improves the performance of the pfr, FLCRM, and FLCRM-F models across all metrics. For the concordance indices, the FLCRM-F model consistently demonstrates higher values than both FLCRM and pfr, indicating better predictive accuracy in both in-sample and out-of-sample predictions. This improvement in FLCRM-F is especially pronounced at higher frailty variances ($\varphi$), where pfr and FLCRM fall short in capturing unobserved heterogeneity. For example, at $\tau = 0.2$ and $\varphi = 2$ with $n = 100$, FLCRM-F achieves a $\text{CI}_{in}$ of 0.878, while pfr and FLCRM reach only 0.785 and 0.790, respectively. These differences emphasize FLCRM-F’s robustness in high-frailty settings.

For both MSE and IMSE, FLCRM-F consistently exhibits the lowest errors, highlighting its effectiveness in estimating both scalar and functional coefficients more accurately than FLCRM and pfr. When $\varphi = 0.01$ and $\tau = 0.01$, pfr’s MSE decreases from 0.026 at $n = 100$ to 0.001 at $n = 1000$, but FLCRM-F maintains a similar MSE while achieving significantly lower IMSE values (0.086 at $n = 100$ versus 0.151 for pfr). As $\varphi$ increases, both pfr and FLCRM see more substantial rises in MSE and IMSE, while FLCRM-F demonstrates smaller increases, showcasing its resilience to unobserved heterogeneity. At high frailty $\varphi = 0.2$ and moderate sample size ($n = 250$) with low censoring parameter ($\tau = 0.01$), for example, FLCRM-F's IMSE is 0.091, which remains lower than the values for pfr (0.207) and FLCRM (0.274).

In scenarios with minimal frailty effects ($\varphi = 0.01$), one might expect similar performance across all methods. However, Table~\ref{tab:tab_1} reveals that FLCRM-F achieves improved estimates compared to both pfr and FLCRM, likely due to its more effective criterion for selecting the number of FPCs based on an explained variance threshold (85\%), as opposed to the AIC criterion used in FLCRM and pfr. This advantage allows FLCRM-F to capture functional effects more precisely, achieving lower IMSE values compared to the other methods even when frailty is low.

With increasing frailty variance ($\varphi$), FLCRM-F continues to outperform both pfr and FLCRM in terms of concordance indices, MSE, and IMSE. The performance gap widens particularly for high frailty and censoring rates. For instance, at $\varphi = 2$, $\tau = 0.2$ with $n = 500$, FLCRM-F maintains a low IMSE of 0.062, while pfr and FLCRM display higher IMSEs of 0.168 and 0.192, respectively. This stability indicates FLCRM-F’s superior ability to account for unobserved heterogeneity, which is less effectively handled by FLCRM and pfr under these challenging conditions.

The observed censoring rate $\psi$ naturally increases with the censoring parameter $\tau$. Across all configurations of $\varphi$ and $n$, the observed censoring rate increases as $\tau$ increases, confirming that the simulation design is working as expected. Higher censoring rates introduce greater difficulty for all the models, but the FLCRM-F model consistently shows superior performance in terms of both concordance indices and error metrics.

Our results indicate that the censoring rate, controlled by the parameter $\tau$, does not significantly impact model performance. As the censoring rate increases from $\tau = 0.01$ to $\tau = 0.2$, the concordance indices, MSE, and IMSE values remain relatively stable for all the models. Additionally, the MSE and IMSE values for all the methods exhibit negligible changes across different values of $\tau$. This outcome may be attributed to the sample sizes used in our simulations and/or the fact that censoring is applied randomly.

An interesting result obtained from Table~\ref{tab:tab_1} is that, the proposed FLCRM-F model consistently outperforms the traditional FLCRM method in both in-sample ($\text{CI}_{in}$) and out-of-sample ($\text{CI}_{out}$) concordance indices when the frailty variance is high. Additionally, FLCRM-F surpasses the pfr method in terms of $\text{CI}_{in}$ under high frailty conditions, although $\text{CI}_{out}$ values between FLCRM-F and pfr remain similar. Importantly, FLCRM-F demonstrates stronger performance in parameter estimation accuracy compared to pfr, further validating its robustness when frailty effects are present. The observed similarity in $\text{CI}_{out}$ values between FLCRM-F and pfr, despite FLCRM-F’s superior parameter estimation, likely stems from the nature of the concordance index as a metric. The concordance index measures the rank correlation between predicted risk scores and observed survival times, focusing primarily on the order of events rather than the exact magnitude of risk estimates. Consequently, when unobserved heterogeneity (frailty) is high, both FLCRM-F and pfr may achieve similar $\text{CI}_{out}$ values if they correctly capture event ordering. However, FLCRM-F’s improved parameter estimation reflects a more accurate representation of underlying frailty effects, suggesting that the concordance index alone may not fully capture the enhanced model fit provided by FLCRM-F under frailty.

In summary, while all models benefit from larger sample sizes, the FLCRM-F model stands out with its consistently lower MSE and IMSE, as well as superior predictive accuracy in both in-sample and out-of-sample contexts. These results demonstrate that FLCRM-F is better suited to handle complex survival data involving high frailty variance and censoring rates, making it a robust choice compared to both FLCRM and pfr.

\begin{small}
\begin{center}
\tabcolsep 0.06in
\begin{longtable}{@{}lllccccccccccccc@{}} 
\caption{Point estimates of various model performances considered in our simulation exercise for different settings of samples sizes ($n$), frailty variance ($\varphi$), and level of censoring $(\tau)$. We consider the following model performance metrics: prediction concordance index based on the training set ($\text{CI}_{in}$), out-of-sample prediction concordance index ($\text{CI}_{out}$), Mean squared error ($\text{MSE}$), Integrated mean squared error (IMSE), and estimated censoring level ($\psi$). The estimates are obtained over 500 simulation runs for each settings.}\label{tab:tab_1} \\
\toprule
{$\tau$} & {$\varphi$} & {$n$} & \multicolumn{4}{c}{pfr} & \multicolumn{4}{c}{FLCRM} & \multicolumn{4}{c}{FLCRM-F} & {$\psi$} \\
\cmidrule(l){4-15} \cmidrule(l){16-16} 
&  &  & $\text{CI}_{in}$ & $\text{CI}_{out}$ & $\text{MSE}$ & IMSE & $\text{CI}_{in}$ & $\text{CI}_{out}$ & $\text{MSE}$ & IMSE & $\text{CI}_{in}$ & $\text{CI}_{out}$ & $\text{MSE}$ & IMSE \\
\cmidrule(l){4-16}
\endfirsthead
\toprule
{$\tau$} & {$\varphi$} & {$n$} & \multicolumn{4}{c}{pfr} & \multicolumn{4}{c}{FLCRM} & \multicolumn{4}{c}{FLCRM-F} & {$\psi$}  \\
\cmidrule(l){4-16}
&  &  & $\text{CI}_{in}$ & $\text{CI}_{out}$ & $\text{MSE}$ & IMSE & $\text{CI}_{in}$ & $\text{CI}_{out}$ & $\text{MSE}$ & IMSE & $\text{CI}_{in}$ & $\text{CI}_{out}$ & $\text{MSE}$ & IMSE \\
\cmidrule(l){4-16}
\endhead
\cmidrule(l){3-16}
\multicolumn{16}{r}{Continued on next page} \\ 
\endfoot
\endlastfoot
\multirow{16}{*}{0.01} & \multirow{4}{*}{0.01} & 100 & 0.843 & 0.825 & 0.026 & 0.151 & 0.847 & 0.755 & 0.031 & 0.282 & 0.852 & 0.828 & 0.034 & 0.086 & 0.051  \\
& & 250 & 0.840 & 0.834 & 0.008 & 0.087 & 0.842 & 0.763 & 0.009 & 0.079 & 0.846 & 0.835 & 0.009 & 0.028 & 0.050  \\
& & 500 & 0.839 & 0.835 & 0.004 & 0.064 & 0.840 & 0.758 & 0.004 & 0.040 & 0.843 & 0.836 & 0.004 & 0.016 & 0.050 \\
& & 1000 & 0.839 & 0.837 & 0.001 & 0.052 & 0.839 & 0.756 & 0.001 & 0.021 & 0.842 & 0.837 & 0.002 & 0.010 & 0.050  \\
\cmidrule(l){4-16}
& \multirow{4}{*}{1} & 100 & 0.792 & 0.769 & 0.069 & 0.182 & 0.797 & 0.713 & 0.064 & 0.458 & 0.873 & 0.772 & 0.051 & 0.129 & 0.093  \\
& & 250 & 0.787 & 0.776 & 0.057 & 0.153 & 0.790 & 0.719 & 0.054 & 0.196 & 0.888 & 0.781 & 0.023 & 0.053 & 0.097  \\
& & 500 & 0.787 & 0.782 & 0.054 & 0.135 & 0.788 & 0.719 & 0.052 & 0.152 & 0.894 & 0.785 & 0.014 & 0.032 & 0.095 \\
& & 1000 & 0.787 & 0.784 & 0.053 & 0.120 & 0.787 & 0.728 & 0.053 & 0.120 & 0.896 & 0.786 & 0.010 & 0.022 & 0.095  \\
\cmidrule(l){4-16}
& \multirow{4}{*}{1.5} & 100 & 0.772 & 0.744 & 0.090 & 0.208 & 0.776 & 0.690 & 0.093 & 0.496 & 0.880 & 0.747 & 0.070 & 0.143 & 0.128  \\
& & 250 & 0.764 & 0.753 & 0.083 & 0.188 & 0.767 & 0.701 & 0.079 & 0.240 & 0.899 & 0.758 & 0.034 & 0.070 & 0.129  \\
& & 500 & 0.764 & 0.758 & 0.079 & 0.173 & 0.766 & 0.707 & 0.078 & 0.189 & 0.904 & 0.761 & 0.025 & 0.047 & 0.130 \\
& & 1000 & 0.764 & 0.761 & 0.079 & 0.158 & 0.765 & 0.703 & 0.078 & 0.156 & 0.907 & 0.762 & 0.020 & 0.036 & 0.128  \\
\cmidrule(l){4-16}
& \multirow{4}{*}{2} & 100 & 0.752 & 0.719 & 0.112 & 0.241 & 0.756 & 0.675 & 0.112 & 0.494 & 0.880 & 0.725 & 0.087 & 0.166 & 0.168  \\
& & 250 & 0.746 & 0.733 & 0.100 & 0.207 & 0.749 & 0.681 & 0.099 & 0.274 & 0.902 & 0.735 & 0.049 & 0.091 & 0.164  \\
& & 500 & 0.744 & 0.737 & 0.098 & 0.202 & 0.747 & 0.690 & 0.096 & 0.206 & 0.907 & 0.742 & 0.037 & 0.069 & 0.167 \\
& & 1000 & 0.745 & 0.741 & 0.097 & 0.190 & 0.745 & 0.692 & 0.097 & 0.185 & 0.911 & 0.744 & 0.032 & 0.057 & 0.164  \\
\cmidrule(l){2-16}
\multirow{16}{*}{0.1} & \multirow{4}{*}{0.01} & 100 & 0.849 & 0.831 & 0.034 & 0.159 & 0.853 & 0.759 & 0.038 & 0.332 & 0.858 & 0.835 & 0.040 & 0.098 & 0.203  \\
& & 250 & 0.846 & 0.840 & 0.010 & 0.097 & 0.847 & 0.763 & 0.011 & 0.085 & 0.851 & 0.840 & 0.012 & 0.032 & 0.204  \\
& & 500 & 0.845 & 0.841 & 0.004 & 0.070 & 0.844 & 0.761 & 0.004 & 0.041 & 0.848 & 0.842 & 0.005 & 0.017 & 0.202 \\
& & 1000 & 0.844 & 0.843 & 0.002 & 0.054 & 0.844 & 0.765 & 0.002 & 0.023 & 0.847 & 0.843 & 0.002 & 0.011 & 0.203  \\
\cmidrule(l){4-16}
& \multirow{4}{*}{1} & 100 & 0.805 & 0.782 & 0.068 & 0.181 & 0.808 & 0.725 & 0.067 & 0.419 & 0.870 & 0.785 & 0.060 & 0.142 & 0.283  \\
& & 250 & 0.800 & 0.791 & 0.048 & 0.146 & 0.802 & 0.729 & 0.046 & 0.190 & 0.886 & 0.794 & 0.025 & 0.055 & 0.282  \\
& & 500 & 0.800 & 0.794 & 0.044 & 0.128 & 0.801 & 0.733 & 0.043 & 0.139 & 0.894 & 0.797 & 0.014 & 0.031 & 0.281 \\
& & 1000 & 0.800 & 0.797 & 0.043 & 0.106 & 0.801 & 0.735 & 0.042 & 0.106 & 0.897 & 0.798 & 0.009 & 0.022 & 0.281  \\
\cmidrule(l){4-16}
& \multirow{4}{*}{1.5} & 100 & 0.789 & 0.761 & 0.086 & 0.202 & 0.796 & 0.705 & 0.086 & 0.602 & 0.877 & 0.763 & 0.073 & 0.153 & 0.325  \\
& & 250 & 0.783 & 0.771 & 0.067 & 0.175 & 0.786 & 0.715 & 0.065 & 0.239 & 0.897 & 0.774 & 0.034 & 0.070 & 0.323  \\
& & 500 & 0.781 & 0.774 & 0.064 & 0.156 & 0.783 & 0.719 & 0.062 & 0.164 & 0.904 & 0.778 & 0.023 & 0.048 & 0.324 \\
& & 1000 & 0.782 & 0.778 & 0.063 & 0.140 & 0.782 & 0.724 & 0.062 & 0.134 & 0.905 & 0.780 & 0.018 & 0.035 & 0.323  \\
\cmidrule(l){4-16}
& \multirow{4}{*}{2} & 100 & 0.773 & 0.740 & 0.105 & 0.214 & 0.780 & 0.688 & 0.104 & 0.575 & 0.874 & 0.742 & 0.090 & 0.179 & 0.365  \\
& & 250 & 0.768 & 0.753 & 0.084 & 0.190 & 0.771 & 0.704 & 0.081 & 0.250 & 0.901 & 0.757 & 0.047 & 0.092 & 0.365  \\
& & 500 & 0.766 & 0.759 & 0.080 & 0.179 & 0.767 & 0.700 & 0.079 & 0.186 & 0.908 & 0.763 & 0.035 & 0.063 & 0.363 \\
& & 1000 & 0.766 & 0.761 & 0.078 & 0.164 & 0.767 & 0.707 & 0.078 & 0.153 & 0.911 & 0.764 & 0.029 & 0.050 & 0.362  \\
\cmidrule(l){2-16}
\multirow{16}{*}{0.2} & \multirow{4}{*}{0.01} & 100 & 0.852 & 0.835 & 0.036 & 0.168 & 0.856 & 0.764 & 0.042 & 0.297 & 0.862 & 0.838 & 0.044 & 0.105 & 0.281  \\
& & 250 & 0.849 & 0.842 & 0.011 & 0.100 & 0.850 & 0.765 & 0.012 & 0.094 & 0.855 & 0.844 & 0.013 & 0.035 & 0.281  \\
& & 500 & 0.848 & 0.844 & 0.005 & 0.072 & 0.848 & 0.773 & 0.005 & 0.048 & 0.852 & 0.845 & 0.005 & 0.018 & 0.281 \\
& & 1000 & 0.848 & 0.845 & 0.002 & 0.057 & 0.848 & 0.765 & 0.002 & 0.025 & 0.851 & 0.846 & 0.002 & 0.011 & 0.281  \\
\cmidrule(l){4-16}
& \multirow{4}{*}{1} & 100 & 0.814 & 0.786 & 0.070 & 0.181 & 0.818 & 0.732 & 0.070 & 0.546 & 0.870 & 0.789 & 0.066 & 0.147 & 0.363  \\
& & 250 & 0.809 & 0.796 & 0.046 & 0.143 & 0.811 & 0.732 & 0.043 & 0.214 & 0.891 & 0.800 & 0.025 & 0.058 & 0.361  \\
& & 500 & 0.807 & 0.801 & 0.040 & 0.124 & 0.808 & 0.735 & 0.039 & 0.147 & 0.894 & 0.803 & 0.015 & 0.033 & 0.360 \\
& & 1000 & 0.807 & 0.804 & 0.038 & 0.105 & 0.807 & 0.738 & 0.037 & 0.100 & 0.898 & 0.805 & 0.009 & 0.022 & 0.362  \\
\cmidrule(l){4-16}
& \multirow{4}{*}{1.5} & 100 & 0.797 & 0.765 & 0.087 & 0.204 & 0.801 & 0.714 & 0.088 & 0.488 & 0.878 & 0.768 & 0.078 & 0.168 & 0.406  \\
& & 250 & 0.790 & 0.779 & 0.064 & 0.169 & 0.794 & 0.720 & 0.062 & 0.226 & 0.898 & 0.781 & 0.036 & 0.079 & 0.401  \\
& & 500 & 0.789 & 0.783 & 0.059 & 0.153 & 0.791 & 0.726 & 0.058 & 0.156 & 0.904 & 0.786 & 0.024 & 0.046 & 0.403 \\
& & 1000 & 0.790 & 0.786 & 0.057 & 0.133 & 0.790 & 0.731 & 0.056 & 0.132 & 0.908 & 0.788 & 0.017 & 0.035 & 0.403  \\
\cmidrule(l){4-16}
& \multirow{4}{*}{2} & 100 & 0.785 & 0.750 & 0.103 & 0.212 & 0.790 & 0.693 & 0.102 & 0.651 & 0.878 & 0.752 & 0.089 & 0.193 & 0.441  \\
& & 250 & 0.777 & 0.763 & 0.078 & 0.186 & 0.778 & 0.711 & 0.076 & 0.269 & 0.901 & 0.767 & 0.045 & 0.091 & 0.439  \\
& & 500 & 0.776 & 0.767 & 0.074 & 0.168 & 0.778 & 0.714 & 0.071 & 0.192 & 0.908 & 0.771 & 0.033 & 0.062 & 0.439 \\
& & 1000 & 0.776 & 0.772 & 0.071 & 0.156 & 0.777 & 0.718 & 0.071 & 0.152 & 0.912 & 0.774 & 0.028 & 0.048 & 0.440  \\
\bottomrule
\end{longtable}
\end{center}
\end{small}

\begin{figure}[!htb]
\centering
\includegraphics[width=5.75cm]{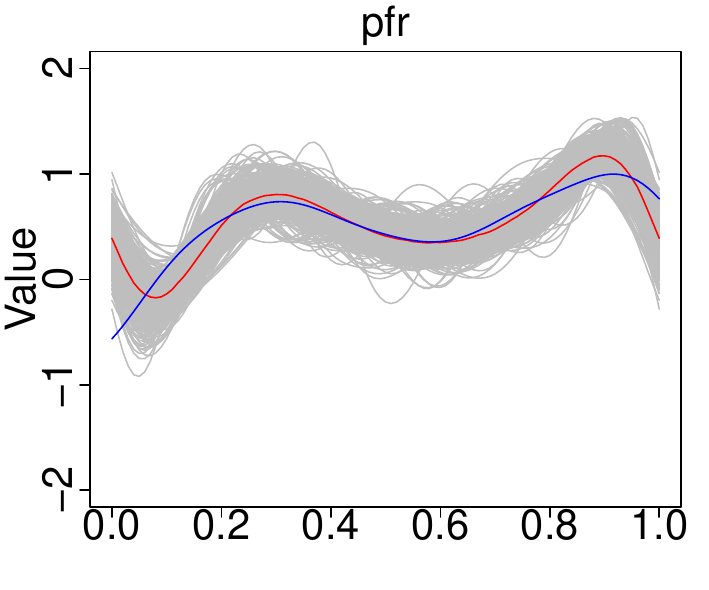}
\quad
\includegraphics[width=5.75cm]{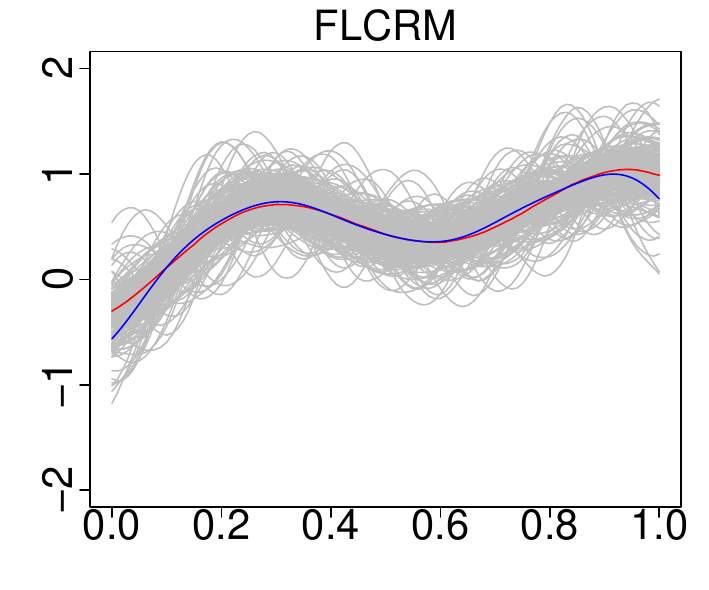}
\quad
\includegraphics[width=5.75cm]{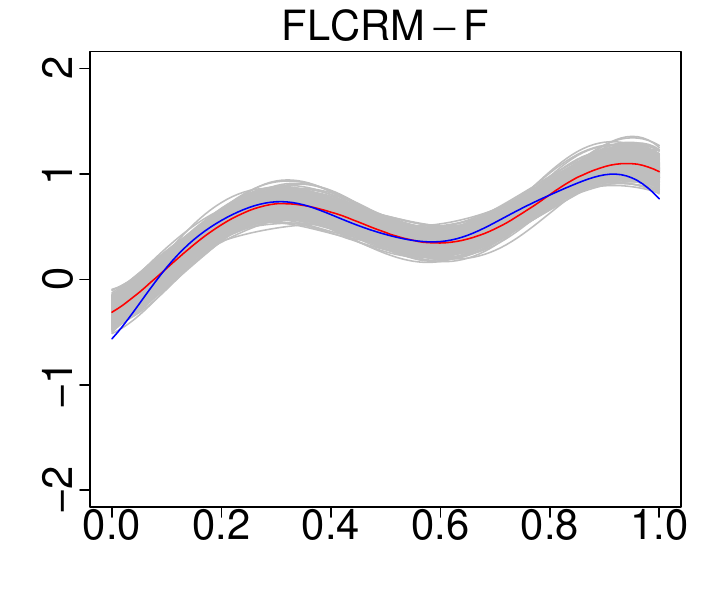}
\\ 
\includegraphics[width=5.75cm]{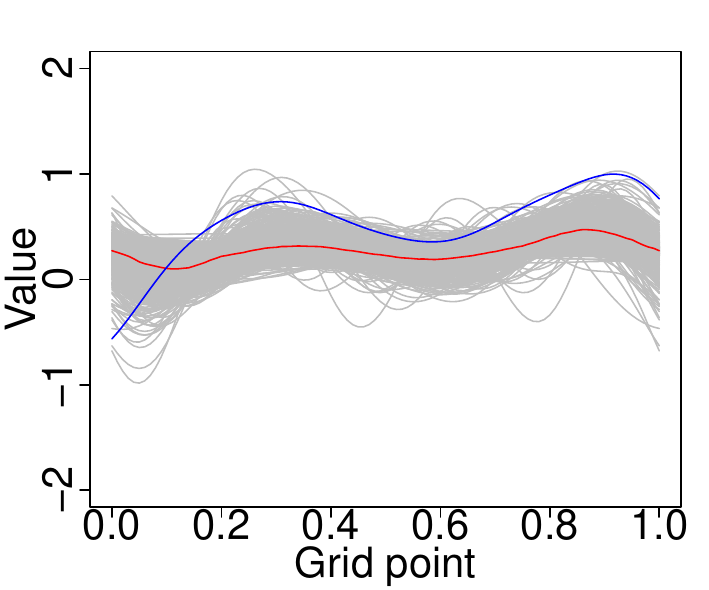}
\quad
\includegraphics[width=5.75cm]{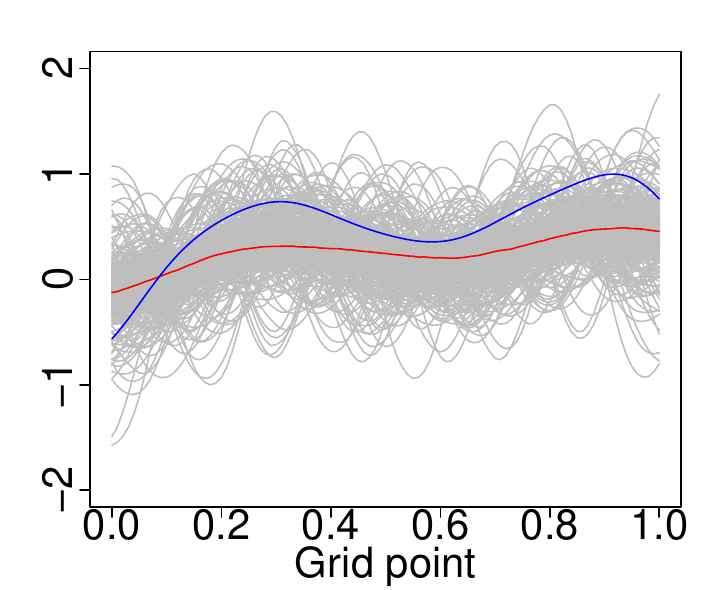}
\quad
\includegraphics[width=5.75cm]{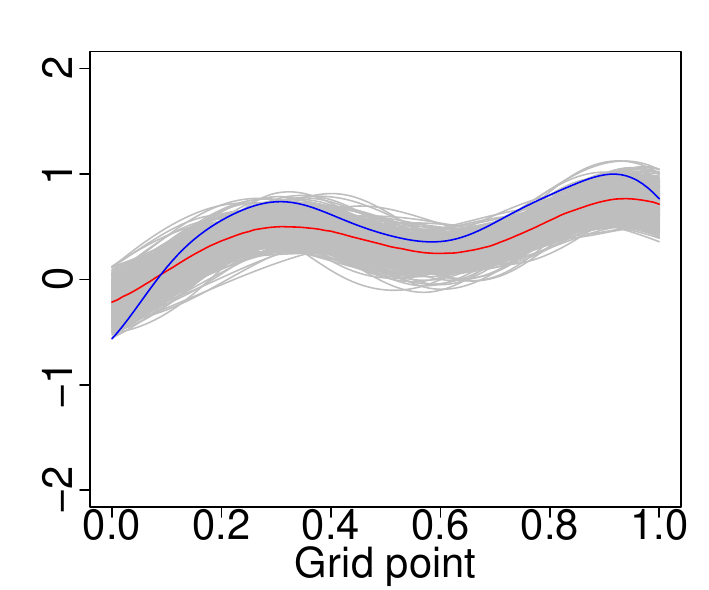}
\caption{\small{Graphical displays of the parameter estimates (grey curves), their means (red curves), and true parameter functions (blue curves) for the pfr (left panels), FLCRM (middle panels), and FLCRM-F (right panels) when the sample size is $n = 1000$. The results obtained for $\tau = 0.01$ and $\varphi = 0.01$ are shown in the first row, while the results for $\tau = 0.2$ and $\varphi =2$ are presented in the second row.}.}\label{fig:Fig_2}
\end{figure}

Figure~\ref{fig:Fig_2} presents graphical illustrations of the parameter estimates derived from 500 Monte Carlo simulations, alongside their corresponding means and the true parameter functions. The visual results further corroborate the conclusions drawn from Table~\ref{tab:tab_1}, demonstrating that the proposed method yields consistently stable parameter estimates closely aligned with the true parameter functions across all values of $\tau$ and $\varphi$.

\section{Application}\label{sec:5}

Our application is based on the 2011-2012 and 2013-2014 National Health and Nutrition Examination Survey (NHANES), which provides a comprehensive view of health-related metrics for a nationally representative sample of the U.S. population. Conducted by the National Center for Health Statistics (NCHS), a branch of the Centers for Disease Control and Prevention (CDC), NHANES combines both interview and examination components. Each year, the survey samples approximately 5,000 individuals from diverse demographic and socioeconomic backgrounds, capturing key information about health, lifestyle, and nutrition. Survey participants respond to a range of questions covering demographic characteristics, socioeconomic factors, diet, and various health-related issues, supplemented by a rigorous examination component that includes physical measurements and laboratory assessments.

For the analysis, we focus on a subset of NHANES participants aged 50 to 80 (3520 subjects in total), utilizing single-level data that includes both functional and scalar variables. The key functional variable, Monitor-Independent Movement summaries(MIMS) \citep{john2019open}, is derived from the Physical Activity Monitor (PAM) component. Participants wore a triaxial accelerometer for 24 hours over seven consecutive days, capturing minute-by-minute physical activity data. For each participant, the data was averaged across available days, resulting in 1,440 observations per participant, corresponding to the total minutes in a 24-hour day. This high-resolution functional data allows for a nuanced view of physical activity patterns, revealing not only overall activity levels but also the variation in intensity and duration of movement throughout the day and week. By modeling MIMS as a functional predictor, the data enables in-depth exploration of physical activity’s role in health and disease.

In addition to the functional variable, the analysis includes several scalar variables essential for understanding individual differences in health outcomes. Age, body mass index (BMI), gender, coronary heart disease (CHD) status, and race (the factor variable race has six levels: Mexican American, Non-Hispanic White, Non-Hispanic Black, Non-Hispanic Asian, Other Hispanic, Other Race) are used as scalar covariates, offering valuable contextual information about participants. Age and BMI provide basic demographic and health status markers, while gender, race, and CHD status allow for examining potential gender-specific and race-specific health patterns and the impact of pre-existing cardiovascular conditions on mortality and physical activity.

This NHANES dataset subset, organized by \cite{Crainiceanu2024} and accessed through their website (\url{https://functionaldataanalysis.org/dataset_nhanes.html}), also includes time-to-death data with censoring information. This survival component allows for a comprehensive examination of how physical activity patterns, alongside demographic and health variables, contribute to mortality risk and long-term health trajectories. By combining MIMS with age, BMI, gender, CHD status, and race, the model enables a holistic analysis of health determinants, shedding light on complex interactions between physical activity and mortality among different population groups.

Because NHANES uses a complex, multistage, and probability sampling design to select participants, sample weights were adjusted for in our analyses to ensure that the data is representative of the U.S. population and reduce the estimation biases. The sample weights were generated in three steps: 1) Base weight was first calculated for each sample unit to account for the oversampling of certain population subgroups; 2) Adjustment was made for non-responses; 3) Post-stratification adjustment was made to
make sure the samples match known population totals for demographic characteristics such as age, gender, and race/ethnicity, as provided by the \cite{cdc_weighting}, also see \cite{chen2024antioxidants}.

\begin{figure}[!htb]
\centering
\includegraphics[width=5.8cm]{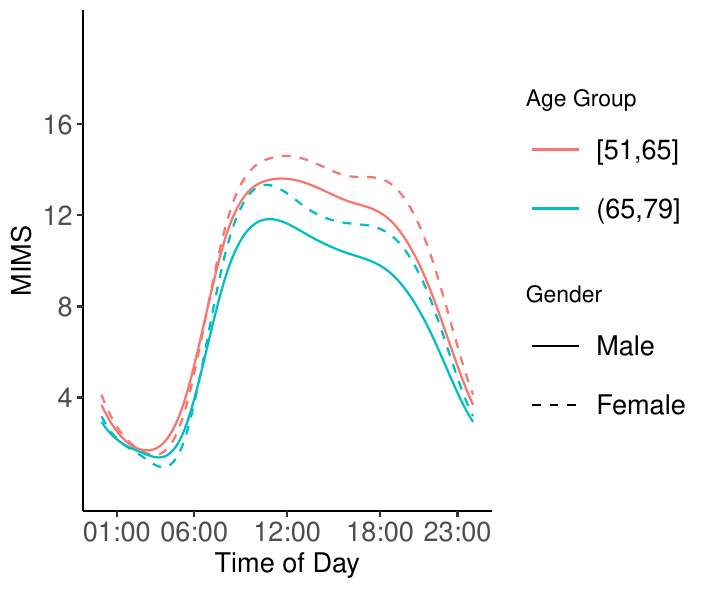}
\quad
\includegraphics[width=5.8cm]{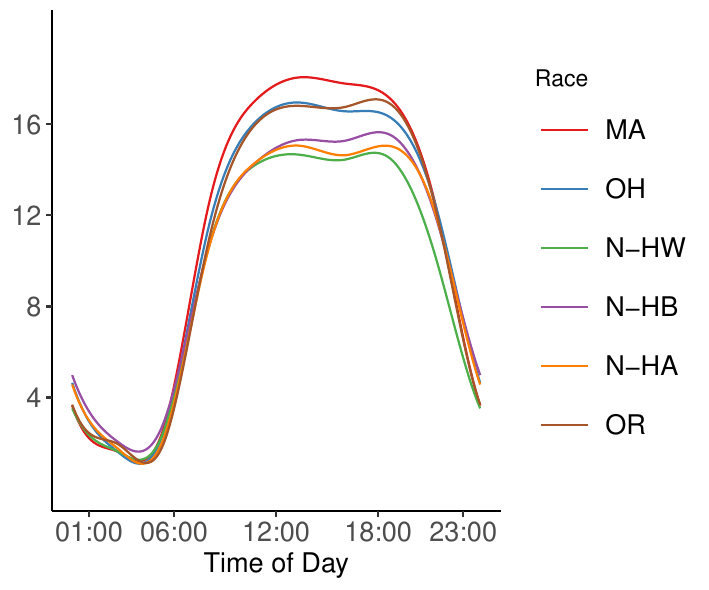}
\quad
\includegraphics[width=5.8cm]{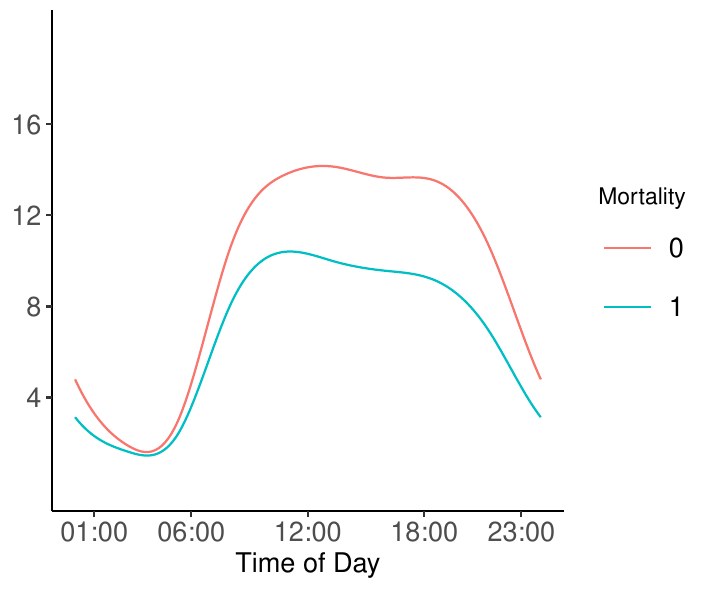}
\caption{\small{A graphical display of daily physical activity patterns segmented by demographic and health status indicators: The mean MIMS values stratified by age group and gender over hourly ntervals (left panel), mean MIMS levels across race (middle panel; ``MA'': Mexican American, ``OH'': Other Hispanic, ``N-HW'': Non-Hispanic White, ``N-HB'': Non-Hispanic Black, ``N-HA'': Non-Hispanic Asian, ``OR'': Other Race), and the mean MIMS levels across two health outcome categories—individuals who are alive versus decease (right panel)}.}\label{fig:Fig_M}
\end{figure}

A graphical display of daily physical activity patterns segmented by demographic, racial, and health status indicators is presented in Figure~\ref{fig:Fig_M}. The plot in the left panel illustrates the mean MIMS values stratified by age group and gender over hourly intervals throughout the day. This visualization reveals distinct diurnal activity fluctuations, with peaks typically occurring in the morning and evening hours, and highlights significant variations between age and gender groups. The middle panel presents the mean MIMS levels across six racial categories. This comparison shows notable differences in daily physical activity patterns among racial groups, with certain groups displaying distinct peaks and troughs in activity levels, suggesting potential disparities in physical activity behavior across races. Finally, the right panel contrasts mean MIMS levels across two health outcome categories—individuals who are alive versus deceased. This comparison indicates a consistently lower activity profile for the deceased group, underscoring potential associations between daily activity patterns and mortality risk. Together, these visualizations provide insights into temporal activity trends, racial disparities, and health-related differences in physical activity, forming a basis for further in-depth analysis of MIMS data within the NHANES cohort.

In our analysis with the NHANES dataset, we consider the following FLCRM-F model:
\begin{equation*}
h_i(t \vert \bm{Z}_i, \X_i, \bm{w}) = h_0(t) \exp \left(\bm{Z}_i^\top \bm{\gamma} + \int_0^{24} \X_i(s) \beta(s) ds + \bm{U}_i^\top \bm{w} \right),
\end{equation*}
where, for $i = 1, \ldots, 3520$, $\bm{Z}_i \in \mathbb{R}^5$ includes the age, BMI, gender, CHD, and race values for the $i^\textsuperscript{th}$ subject, and $\X_i(s)$ is the MIMS curve. Initially, we apply our proposed method to the complete dataset, projecting the MIMS curves onto a finite-dimensional space based on the 85\% explained variance criterion. Our preliminary analysis yields a frailty variance of 0.173, indicating a relatively low frailty effect within the NHANES sample.

To provide a more nuanced estimation and predictive evaluation of our method using the NHANES data, we examine two scenarios, which we will refer to as Case-I and Case-II, respectively. In Case-I, we analyze the original dataset, while in Case-II, we modify the data to simulate a higher frailty effect. To amplify frailty characteristics in the current case, we assign an augmented frailty score based on age, BMI, pre-existing health conditions, and race. Specifically, individuals aged over 65, with extreme BMI values (above 40 or below 18.5), or diagnosed with CHD receive significant frailty points, with CHD contributing the highest score increment due to its strong association with increased cardiovascular risks. In addition, the frailty score incorporates race-specific adjustments, reflecting documented health disparities among different racial and ethnic groups. For example, non-Hispanic Black and Mexican American individuals are assigned additional points to account for systemic factors and socioeconomic challenges influencing health outcomes. This enhanced scoring framework integrates the cumulative effects of age, BMI, CHD, and race, constructing a comprehensive, multidimensional measure of health vulnerability that captures both individual and demographic risk factors.

To simulate the influence of frailty on survival, we adjust the time-to-event variable, reducing survival times for participants with elevated frailty scores. For individuals scoring 5 or more, survival times are decreased by 60\%, and for scores of 7 or higher, by 80\%. Additionally, BMI and age are further modified in high-risk groups: participants with BMIs exceeding 40 or below 18.5 have their values adjusted to reflect increased health risks linked to extreme body compositions. These adjustments result in a dataset that models the elevated risk of adverse outcomes in frail individuals, allowing us to assess the interactions between physical activity, frailty, and mortality.

In both scenarios, we randomly divide the dataset into equal-sized training and test sets. Using the training set, we implement the pfr, FLCRM, and FLCRM-F models, calculating $\text{CI}_{in}$ values to assess model fitting performance. Subsequently, these models are applied to the test set, where we compute $\text{CI}_{out}$ to evaluate out-of-sample predictive performance. This entire procedure is repeated 500 times, with results recorded for each iteration, providing a robust assessment of model performance under varying frailty conditions.

Our findings are presented in Table~\ref{tab:tab_3}. For Case-I, the FLCRM-F model achieves a considerably higher mean $\text{CI}_{in}$ value (0.915) compared to both the pfr and FLCRM models, which each report a $\text{CI}_{in}$ of 0.735. This result suggests an improved model fit for FLCRM-F when no additional frailty effect is introduced. The $\text{CI}_{out}$ values further highlight this advantage, with the FLCRM-F model matching the performance of the pfr model at 0.745 but significantly outperforming the FLCRM model, which shows a sharp decline to 0.606, indicating poor out-of-sample predictive capacity for FLCRM under standard conditions.

In Case-II, where the dataset was modified to incorporate a higher frailty effect, the FLCRM-F model’s performance is even more pronounced. It reaches the highest $\text{CI}_{in}$ (0.960) and $\text{CI}_{out}$ (0.859) values, indicating robust model fitting and predictive capability even with an increased frailty effect. The FLCRM model, while showing some improvement in $\text{CI}_{out}$ (0.741) relative to Case-I, remains substantially lower than FLCRM-F, highlighting the limitations of FLCRM under heightened frailty. The pfr method produces improved performance compared to its results obtained under Case-I but is outperformed by FLCRM-F in each scenario. Overall, these results underscore the effectiveness of FLCRM-F in capturing frailty effects and achieving superior model fit and predictive performance, particularly in datasets where frailty significantly influences survival outcomes.

\begin{table}[!htb]
\centering
\tabcolsep 0.35in
\caption{\small{Computed mean $\text{CI}_{in}$ and $\text{CI}_{out}$ values along with their standar errors (in brackets) from the NHANES data.}}\label{tab:tab_3}
\begin{tabular}{@{}lccccc@{}} 
\toprule
{Case} & {Method} & pfr & FLCRM & FLCRM-F \\
\midrule
\multirow{2}{*}{I} & $\text{CI}_{in}$ & 0.735 (0.004) & 0.735 (0.004) &  0.915 (0.095) \\
& $\text{CI}_{out}$ & 0.742 (0.038) & 0.606 (0.128) & 0.745 (0.039) \\
\midrule
\multirow{2}{*}{II} & $\text{CI}_{in}$ & 0.824 (0.003) & 0.823 (0.003) &  0.960 (0.057) \\
& $\text{CI}_{out}$ & 0.825 (0.032) & 0.741 (0.078) & 0.859 (0.031) \\
\bottomrule
\end{tabular}
\end{table}

Using the full NHANES dataset, we applied the pfr, FLCRM, and FLCRM-F methods to estimate the scalar parameters $\widehat{\gamma}_1$, $\widehat{\gamma}_2$, $\widehat{\gamma}_3$, $\widehat{\gamma}_4$, $\widehat{\gamma}_{5,\text{OH}}$, $\widehat{\gamma}_{5,\text{N-HW}}$, $\widehat{\gamma}_{5,\text{N-HB}}$, $\widehat{\gamma}_{5,\text{N-HA}}$, $\widehat{\gamma}_{5,\text{OR}}$, which correspond to age, BMI, gender, CHD, and the five race categories (with Mexican American as the reference group), respectively. The results are presented in Table~\ref{tab:tab_res}. In Case-I, where the original dataset was used without modifications to frailty, for age ($\widehat{\gamma}_1$), the estimates ranged from 0.056 to 0.091, with FLCRM-F providing the highest value. For BMI ($\widehat{\gamma}_2$),  the estimates were slightly variable, with pfr estimating $-0.015$, FLCRM estimating $0.008$ and FLCRM-F estimating $-0.009$. Gender ($\widehat{\gamma}_3$) estimates showed greater variation, with FLCRM estimating the most negative value ($-0.207$) and pfr the least ($-0.045$). The largest differences were observed in the CHD estimates ($\widehat{\gamma}_4$), where FLCRM-F provided a significantly higher value ($1.401$) compared to pfr ($0.614$) and FLCRM ($0.625$). The race variables have notable differences across racial categories, with pfr and FLCRM showing small to moderate variations, while FLCRM-F demonstrated more pronounced differences, such as a strongly negative estimate for Non-Hispanic White ($-0.931$) and Other Race ($-0.667$).

In Case-II, where the dataset was adjusted to amplify the frailty effect, larger divergences emerged among the methods, particularly in the CHD and race estimates. For $\widehat{\gamma}_1$ (age), all methods showed increases compared to Case-I, with FLCRM-F again estimating the highest value ($0.143$). BMI estimates ($\widehat{\gamma}_2$) also increased, particularly in FLCRM-F ($0.048$). Gender ($\widehat{\gamma}_3$) estimates remained relatively stable across methods, with only slight variation. However, CHD ($\widehat{\gamma}_4$) exhibited the most significant increase, with FLCRM-F estimating a notably higher value ($5.347$) compared to pfr ($2.622$) and FLCRM ($2.592$). The race estimates also demonstrated greater variability in Case-II, with FLCRM-F providing more extreme values, such as $-1.381$ for Non-Hispanic White and $-1.374$ for Other Race, highlighting the amplified impact of frailty on these associations.

\begin{table}[!htb]
\centering
\tabcolsep 0.1in
\caption{\small{Estimated $\widehat{\gamma}$ values by the methods. $[\widehat{\gamma}_1, \widehat{\gamma}_2, \widehat{\gamma}_3, \widehat{\gamma}_4, \widehat{\gamma}_{5,\text{OH}}, \widehat{\gamma}_{5,\text{N-HW}}, \widehat{\gamma}_{5,\text{N-HB}}, \widehat{\gamma}_{5,\text{N-HA}}, \widehat{\gamma}_{5,\text{OR}}]$ denotes the estimated parameter estimates for age, BMI, gender, CHD, and race variables, respectively.}}\label{tab:tab_res}
\begin{tabular}{@{}lccccccccccc@{}} 
\toprule
{Case} & {Method} & {$\widehat{\gamma}_1$} & {$\widehat{\gamma}_2$} & {$\widehat{\gamma}_3$} & {$\widehat{\gamma}_4$} & {$\widehat{\gamma}_{5,\text{OH}}$} & {$\widehat{\gamma}_{5,\text{N-HW}}$} & {$\widehat{\gamma}_{5,\text{N-HB}}$} & {$\widehat{\gamma}_{5,\text{N-HA}}$} & {$\widehat{\gamma}_{5,\text{OR}}$} \\
\midrule
\multirow{3}{*}{I} & pfr & 0.056 & -0.015 &  -0.045 & 0.614 & -0.308 & 0.008 & 0.022 &  -0.439 & 0.074\\
& FLCRM & 0.066 & 0.008 & -0.207 & 0.625 & -0.345 & -0.167 & 0.029 &  -0.320 & -0.263 \\
& FLCRM-F & 0.091 & -0.009 & -0.084 & 1.401 & -0.688 & -0.931 & -0.124 & -0.528 & -0.667  \\
\midrule
\multirow{3}{*}{II} & pfr & 0.080 & 0.014 &  -0.036 & 2.622 & -0.276 & -0.348 & 0.082 & -0.744 & -0.332 \\
& FLCRM & 0.081 & 0.031 & -0.174 & 2.592 & -0.295 & -0.446 & 0.103 & -0.624 & -0.417 \\
& FLCRM-F & 0.143 & 0.048 & -0.057 & 5.347 & -0.476 & -1.381 & 0.075 & -1.135 & -1.374 \\
\bottomrule
\end{tabular}
\end{table}

\begin{figure}[!htb]
\centering
\includegraphics[width=5.75cm]{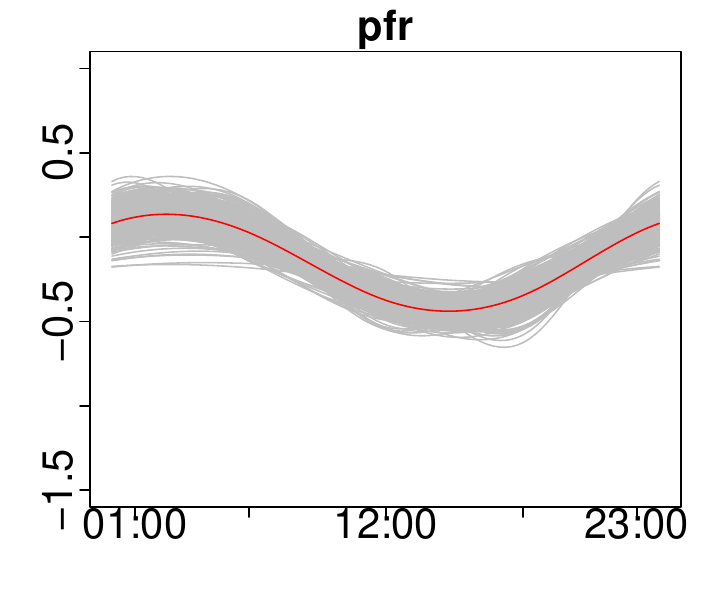}
\quad
\includegraphics[width=5.75cm]{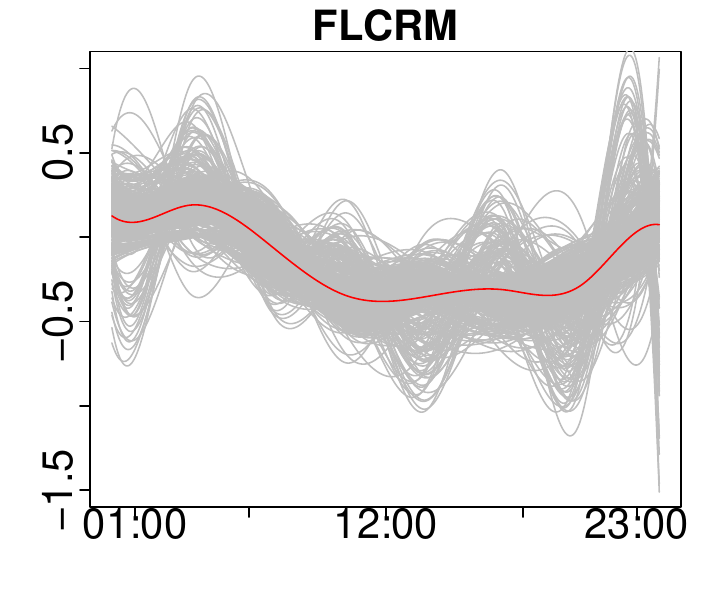}
\quad
\includegraphics[width=5.75cm]{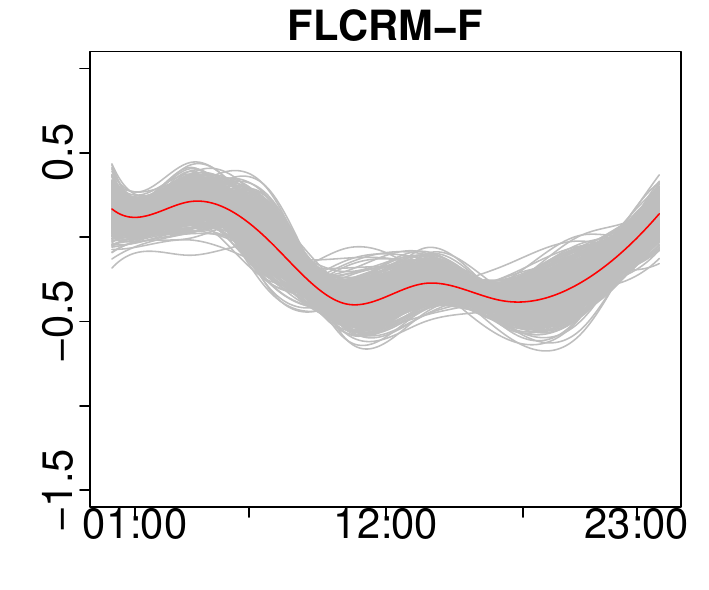}
\\ 
\includegraphics[width=5.75cm]{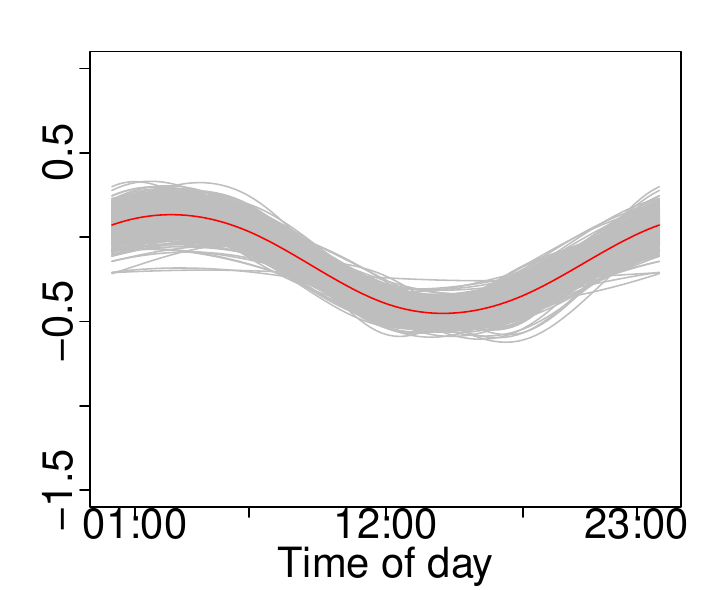}
\quad
\includegraphics[width=5.75cm]{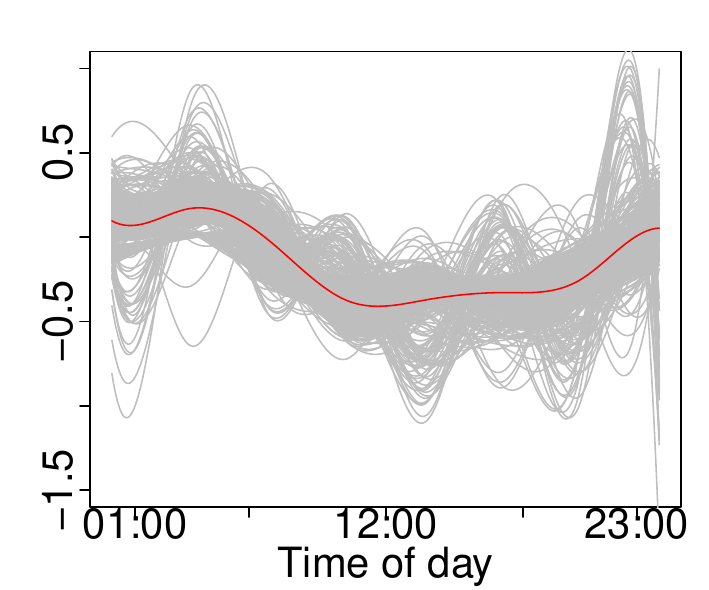}
\quad
\includegraphics[width=5.75cm]{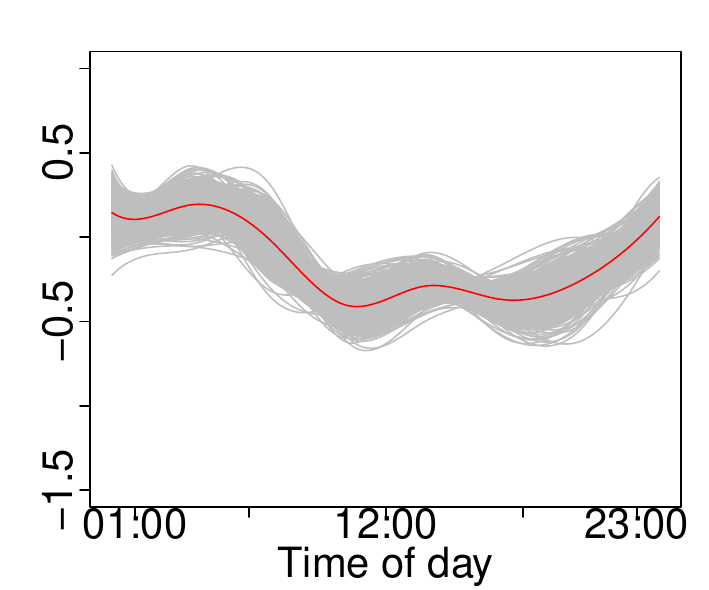}
\caption{\small{Estimated regression coefficient functions (grey curves) and their averages (red curves) for NHANES data. The first rowdisplays results for Case-I, while the second row illustrates outcomes obtained under Case-II. Parameter functions are obtained over 500 bootstrap runs for the pfr (left panels), FLCRM (middle panels), and FCRM-F (right panels) methods (colour figure online)}.}\label{fig:Fig_3}
\end{figure}

To obtain the estimated coefficient functions for each method (pfr, FLCRM, and FLCRM-F), we conducted a bootstrap procedure with 500 repetitions. In each repetition, a resample of 3520 subjects was drawn with replacement from both the original (Case-I) and modified (Case-II) datasets. Each method was then applied to the resampled datasets to estimate the regression coefficient functions, allowing for robust statistical inference. The resulting estimates are displayed in Figure~\ref{fig:Fig_3}. The results indicate that the coefficient estimates remain relatively consistent between Case-I and Case-II for all methods, suggesting robustness across different frailty conditions. The pfr method, leveraging a regularization technique, produces the smoothest coefficient estimates, minimizing fluctuations in the estimated curves. Our proposed FLCRM-F method also produces smooth estimates, though slightly less so than pfr, as it does not use an explicit regularization term. In contrast, FLCRM exhibits considerable waviness in its estimates, showing substantial variability that suggests limited stability in estimating the coefficient functions. These observations highlight the advantage of FLCRM-F in balancing smoothness and adaptability, outperforming FLCRM in terms of stability while effectively capturing key trends in the data.

\section{Conclusion}\label{sec:6}

This study presents a novel extension to the FLCRM by incorporating frailty, developing the FLCRM-F framework to address the complexities of unobserved heterogeneity in survival data with functional covariates. By embedding a frailty term within the FLCRM structure, the proposed model effectively captures individual differences that are not explained by observed covariates, thereby enhancing the robustness and interpretability of survival analyses in settings where functional predictors play a crucial role.

Through rigorous simulation studies, we have demonstrated that FLCRM-F consistently outperforms both the standard FLCRM and the pfr model in estimation accuracy and predictive capability, especially when frailty effects are substantial. The model’s ability to maintain high concordance indices and low estimation errors across varying sample sizes and frailty conditions underscores its resilience to unobserved heterogeneity, making it a valuable tool for applications with complex survival data. Additionally, our analysis of the NHANES data illustrates the practical utility of FLCRM-F in uncovering significant relationships between physical activity patterns, demographic factors, and mortality, particularly within frail subpopulations, providing deeper insights into health dynamics.

The FLCRM-F model addresses a critical gap in the literature by incorporating frailty into functional survival analysis, providing a robust framework for studies that involve both functional covariates and time-to-event data. Future extensions of this framework could include multilevel or correlated frailty structures, offering enhanced flexibility in managing intricate dependencies and heterogeneity within survival data. Additionally, exploring alternative frailty distributions and optimizing computational methods for large-scale applications hold promise for broadening the model's utility across diverse fields. To further improve the estimation and predictive accuracy of FLCRM-F, a penalized version of the model could be developed, incorporating a smoothness penalty to control functional variable behavior. Such an adaptation, akin to the penalized functional regression approach, would enhance both the robustness and precision of parameter estimates, strengthening the model’s applicability and performance in complex survival contexts.

{\it Conflict of Interest}: None declared.

\section*{Funding}
This research was partly supported by the National Institute of Diabetes and Digestive and Kidney Diseases Award numbers 1R01DK132385-01 and R01DK136994-01A1. This research was also partly supported by the Scientific and Technological Research Council of Turkey (TUBITAK) (grant no. 124F096).

\clearpage
\newpage
\bibliographystyle{agsm}
\bibliography{fcoxf.bib}

\end{document}